\documentclass[conference]{IEEEtran}
\IEEEoverridecommandlockouts
\usepackage[T1]{fontenc}
\usepackage{cite}
\usepackage{amsmath,amssymb,amsfonts}
\usepackage{algorithmic}
\usepackage{graphicx}
\usepackage{textcomp}
\usepackage{xcolor}
\def\BibTeX{{\rm B\kern-.05em{\sc i\kern-.025em b}\kern-.08em
    T\kern-.1667em\lower.7ex\hbox{E}\kern-.125emX}}
\begin{document}
\newcommand{\todo}[1]{{\textcolor{red}{[#1]}}}
\newcommand{\olivia}[1]{{\textcolor{magenta}{[#1]}}}
\newcommand{\tim}[1]{{\textcolor{purple}{[#1]}}}
\newcommand{\steve}[1]{{\textcolor{blue}{[#1]}}}
\newcommand{\michal}[1]{{\textcolor{green}{[#1]}}}
\newcommand{\santiago}[1]{{\textcolor{teal}{[#1]}}}

\makeatletter 
\newcommand{\linebreakand}{%
  \end{@IEEEauthorhalign}
  \hfill\mbox{}\par
  \mbox{}\hfill\begin{@IEEEauthorhalign}
}
\makeatother 

\title{An Abstraction Hierarchy Toward Productive Quantum Programming}

\author{
\IEEEauthorblockN{Olivia Di Matteo}
\IEEEauthorblockA{Electrical and Computer Engineering \\
\emph{The University of British Columbia}\\Vancouver, Canada \\olivia@ece.ubc.ca\\
ORCID 0000-0002-1372-7706}
\and
\IEEEauthorblockN{Santiago Núñez-Corrales}
\IEEEauthorblockA{NCSA/IQUIST \\
\emph{University of Illinois Urbana-Champaign}\\Urbana, IL, USA \\nunezco2@illinois.edu\\
ORCID 0000-0003-4342-6223}
\and
\IEEEauthorblockN{Michał Stęchły}
\IEEEauthorblockA{\textit{Musty Tech} \\
Oshawa, ON, CA \\
michal@mustythoughts.com \\
ORCID 0000-0002-2391-9664}
\and 

\linebreakand 

\IEEEauthorblockN{Steven P. Reinhardt}
\IEEEauthorblockA{\textit{Transform Computing, Inc.} \\
Eagan, MN, USA \\
ORCID 0000-0003-4355-6693}
\and
\IEEEauthorblockN{Tim Mattson}
\IEEEauthorblockA{\textit{Human Learning Group} \\
Ocean Park, WA, USA\\
ORCID 0000-0002-6106-8717}
}

\maketitle

\begin{abstract}

Experience from seven decades of classical computing suggests that a sustainable computer industry depends on a community of software engineers  writing programs to address a wide variety of specific end-user needs, achieving both performance and utility in the process. Quantum computing is an emerging technology, and we do not yet have the insight to understand what quantum software tools and practices will best support researchers, software engineers, or applications specialists. Developers for today's quantum computers are grappling with the low-level details of the hardware, and progress towards scalable devices does not yet suggest what higher-level abstractions may look like. In this paper, we analyze and reframe the current state of the quantum software stack using the language of programming models. We propose an abstraction hierarchy to support quantum software engineering and discuss the consequences of overlaps across the programming, execution, and hardware models found in current technologies. We exercise this hierarchy for solving the eigenvalue estimation problem in two ways (a variational algorithm with error mitigation, and phase estimation with error correction) and pinpoint key differences in these approaches in terms of these layered models and their overlaps. While our work points to concrete conceptual challenges and gaps in quantum programming and proposes some specific steps forward, our primary thesis is that progress hinges on thinking about the abstraction hierarchy holistically, and not just about its components.

\end{abstract}


\section{Introduction}
Over the past decade, quantum computing (QC) has shifted from small-scale devices in university research labs to a growing industry with billions of dollars in investment. Multiple quantum computers with over one thousand qubits  were announced in the last year. In parallel, there is a growing community of researchers seeking to \emph{use} these devices productively to solve end-user problems, which necessitates the development of \emph{quantum software}. Today's programming environments for quantum computing expose the low-level features of devices, often requiring programmers to implement algorithms at the level of individual quantum gates. This low level of programming comprises a barrier for exploration and technology adoption for those outside of specialty areas.

The wide variety of architectures, qubit technologies, and inevitable presence of noise in noisy intermediate-scale quantum (NISQ) hardware  means that extracting meaningful performance requires a significant amount of expert-level hand-tuning. This has resulted in a software ecosystem of highly specialized tools with limited interoperability, often written by developers with little (or no) formal training in software engineering. In our view, continuing in this direction will make it challenging to deliver
user-friendly software tools that non-quantum experts can productively use to leverage the technology and write algorithms at higher levels of abstraction. Continuing in this direction will likely accumulate technical debt as well.
We must look beyond just supporting physicists running experiments and focus on how we can build communities of application programmers.
In other words, for quantum computing to reach its full potential, we need ways of thinking about quantum software development that will accelerate its evolution from a \textit{craft}, where scarce super-expert practitioners meld together disparate not-completely-understood techniques,  to a \textit{software engineering discipline}, where a large community of capable software engineers use well-defined and approachable interfaces. 
We believe that the lack of such software is a key factor in the mounting evidence across the quantum community indicating the lack of a viable and sustainable path to productive quantum programming.

The parallel computing community was faced with a similar problem  45 years ago with new classes of machines that lacked a solid, cross-platform infrastructure to support the engineering of application software. The  community eventually figured out how to support a robust software ecosystem, but not until it had, in hindsight, wasted years without a disciplined intellectual framework to guide the work. If such a framework had existed up front, convergence around a core software infrastructure to support software engineering in parallel computing would have happened much faster. We believe the QC community can learn from this experience and be more proactive about defining and implementing the software infrastructure that quantum software engineers need. 

To that end, we propose that an abstraction hierarchy is the appropriate intellectual tool to explore the interlocking roles played by different software layers, and show how impedance mismatches across them manifest as limitations in productivity and performance.  Motivated by existing quantum software efforts, we propose  a concrete  abstraction hierarchy for quantum software development.  
We acknowledge that proposing this hierarchy now may seem premature, as so many aspects of QC are nascent.  Subscribing to Martinis' well-known \emph{definite optimism}  approach to quantum hardware \cite{martinis2020definite}, we take the contrary point of view, which is that the nascent nature of the field demands we make early specific attempts despite the risk of being wrong and having to redirect. 
Besides having an abstraction hierarchy, however, we must also  test it by demonstrating how common tasks performed by quantum software practitioners fit that framework. We have chosen to do so with the eigenvalue estimation problem, as there are multiple algorithmic approaches whose corresponding program implementations vary across hierarchy levels.
Though community understanding of quantum applications is incomplete, nevertheless we  propose a set of fundamental algorithmic motifs for quantum computing applications.  

\section{The Case for Abstraction Hierarchies}

The practice of computer science often comes down to the art of finding the right abstractions.  A good abstraction exposes the features of a system that must be directly addressed, while hiding the rest behind an appropriate interface. When a set of abstractions are gathered together, the result is called a \emph{model}. Humans can claim to understand a topic once they have a conceptual model for it.  

In many cases, it is difficult to construct a single model that bridges between mental-space and the elements defined by a particular target domain.  Consequently, we often work in terms of a layered stack, or \emph{hierarchy}, of models.  These models take us from a mental model of algorithms that solve our problem onto an understanding of how the resulting code executes on a physical system.  By organizing our understanding of a problem in terms of an abstraction hierarchy, we can make progress in a well-organized and coherent manner.

\subsection{What makes an abstraction hierarchy good?}
\label{ssec:good_abstraction_hierarchy}

Our goal is to create an abstraction hierarchy that supports the productivity of humans working across the full software stack as they collaborate to build a useful, end-to-end quantum computing system. These humans comprise domain-specific application developers (e.g., chemists, physicists, and other specialists), and those who write compilers and hardware-adjacent software. To achieve this goal, we must outline the criteria for what makes an abstraction hierarchy "good".

First and foremost, a good model supports a \emph{separation of concerns} between these groups of humans.  An application developer should not \emph{need} to know the models used in writing a compiler, and neither the applications nor compiler people should need to know how to implement the hardware.

A good abstraction hierarchy, while supporting a separation of concerns, in our view should \emph{abstract but not hide} other layers. That is, it must be possible for people to reason about each layer in the hierarchy, and how concerns at that layer impact the others.  For example, an OpenMP programmer doesn't need to know how the MESI cache-coherence protocol is implemented, but they should understand how it causes false sharing and \emph{what they need to do} in an algorithm to avoid that problem.  Another way to state this is that it must be possible to move between the different layers so each person in the stack of models has control over which layer they work within.

A good model is \emph{clear}. The elements of the model are distinct, straightforward to understand, and how they interact is well defined.  We are often taught that software design is a ``top-down'' or ``bottom-up'' process. Expert programmers, however,  bounce between layers in the abstraction hierarchy, working at whichever layer is most productive for the emerging design.  This process of development has been dubbed \emph{opportunistic refinement}~\cite{visser1990more}. 

A good model is \emph{complete}: it captures all of the essential elements of the system with a direct mapping from ``mental-space’’ onto the ``implementation’’ of a system.  A good model is \emph{simple}, but not \emph{too} simple; i.e., the model may be simple, but not so simple that the vital issues that must be exploited when using the model are obscured.

Developing a strong abstraction hierarchy with separation of concerns is a social  as well as technical task.
For an application developer to cede control of  a certain topic to a lower layer, they must have confidence that the lower layer will handle that topic effectively.
If that lower layer does not, the app developer may have made a major mistake, especially if the hierarchy does not effectively support moving deftly between its layers. Thus, definitions of models, and communication and feedback about those definitions among the affected humans, are essential pieces of evolving an abstraction hierarchy.

\subsection{Example: Parallel Computing}
\label{ssec:parallel_computing}

\begin{table*}[htbp]
\caption{Examples of the three-model hierarchy applied to three different programming APIs in classical parallel computing. For OpenMP and MPI, we focus on the more common usages across a wide variety of programs.
}
\label{tab:modelsExample}
\begin{center}
\begin{tabular}{|l|l|l|l|}
\hline
                  &   {OpenMP}   &        {MPI}           & {CUDA} \\   
\hline
\textbf{Programming model}          & Loop-Level Parallelism         & SPMD  & Single Instruction Multiple Thread  \\
     Maps algorithms onto              & Single Program Multiple Data (SPMD)    &  Partitioned Global Address Space                                      &  (SIMT)    \\
     source code              & Task Graph                    &                                              &   \\
\hline
\textbf{Execution Model}            & Fork-Join  & Communicating sequential processes      & SIMT \\
 Defines abstractions for how       & & (CSP)    &     \\
 code executes & & & \\
\hline
 \textbf{Hardware model}           &   Multiprocessor   & Multi-computer     &General Purpose GPU (GPGPU)   \\
 Maps program execution onto & & & \\
 models of computer systems & & & \\
\hline
\end{tabular}
\end{center}
\end{table*}%

The parallel computing community converged around an abstraction hierarchy composed of three layers.
\begin{itemize}
\item \textbf{Programming model}: Maps the mental representation of a problem onto algorithms expressed in code. A programming model, at the most fundamental level, defines syntax to expose the execution model, while making core design patterns easy to express. 
\item \textbf{Execution model}: Defines an abstract representation of program execution on a  mid-level abstract machine.

\item \textbf{Hardware model}: Maps program execution onto models that describe actual computer systems. These low-level models expose the features of hardware that programmers need to reason about to optimize program execution. While programmers may need to be aware of these features in a general manner, the execution level should provide appropriate subordination of detail.

\end{itemize} 

Together, these models take us from a mental model of algorithms that solve our problem onto an understanding of how the resulting code executes on a physical system.   This hierarchy is  not unique.  There are other ways to build systems of abstraction that assist the design of programming environments.  For example, execution and programming models are sometimes merged into a single model.   Leslie Valiant,  in his classic paper on Bulk Synchronous programming (BSP)~\cite{valiant1990bridging}, combined all three models into a single ``bridging model’’.   For low-level programming, programmers may skip the programming model and work directly in the execution model. 
Note that we don't claim that the three layers of parallel processing is universal; we do assert that the layered notion is.

As an example of how all these models fit together, Table~\ref{tab:modelsExample} presents three common parallel programming environments and how they map onto this three-level abstraction hierarchy.   OpenMP~\cite{desupinski2018ongoing} is a popular parallel programming environment. Focusing on the traditional, core semantics (i.e., leaving aside recent constructs for vector-unit and GPU programming), the hardware model used for OpenMP assumes multiple processing elements with a cache-coherent shared address space, the so-called \textit{multiprocessor} hardware model.  The program starts as a single thread, but \textit{forks} threads as needed when opportunities for parallelism are encountered. When parallel execution completes, the threads \textit{join} together and a single thread proceeds.  This execution model supports a number of key design patterns~\cite{mattson2004patterns} such as loop-level parallelism, Single Program Multiple Data (SPMD), and task-graph patterns.

MPI (the Message Passing Interface)~\cite{gropp2015message} is the standard way to write software for the \emph{multi-computer} (or \emph{shared-nothing}) hardware model.  This is a realization of the communicating sequential processes (CSP) execution model. As the name implies, a set of processes execute independently.  As far as the model is concerned, they are sequential and any interaction between them occurs only through distinct communication events.  While the MPI standard lets a programmer directly code at the level of CSP, in most cases the SPMD or partitioned-global-address-space patterns are exploited.  

Finally, for GPUs the General Purpose GPU hardware model is used.  The GPGPU model in a modern GPU is the result of a codesign process where the hardware has been designed around the needs of a particular execution model: the Single Instruction Multiple Thread, or SIMT model. An index space is defined and an instance of a function (or kernel) defines a single stream of instructions that are mapped onto each point in that space. A thread (or, more properly, work-item) executes at each point. While higher-level approaches based on the loop-level parallelism pattern can be used, most GPU programmers work directly in the SIMT execution model. 

This three-level hierarchy  brings order to the chaos of parallel programming environments.   It lets us talk about families of execution models that map onto a small number of hardware models found in actual systems.  Working with this hierarchy, we can describe how common patterns, such as SPMD, map onto radically different execution models. This organization of abstractions was arrived at after the fact; that is, parallel programmers were faced with a plethora of programming environments and struggled to understand how they were related to each other.   It would have greatly helped if the community had this abstraction hierarchy from the beginning, when it was first figuring out how to program these new types of high-performance computing systems. 

\section{An Abstraction Hierarchy for Quantum Programming}
\label{sec:quantum-abstraction-hierarchy}

Writing programs to use quantum computers for real-world problems today is a major challenge. In our view, one of the main issues is the absence of well-established models. Without models, we lack awareness of the boundaries between them that enable the separation of concerns necessary for productive programming. Quantum programmers generally have to work at \emph{all} levels of the stack, from the high-level algorithm description, to gates, to the hardware-adjacent expression of the algorithm. 
We propose to remedy this with the hierarchy in Figure~\ref{fig:quantum-abstract-hierarchy}. 

In this section, much in the spirit of the description contained in Table \ref{tab:modelsExample}, we motivate this hierarchy by concentrating on key factors that explain the current state of affairs and the challenges derived from it. We articulate connections across those levels into a unified view by travelling upwards from the hardware model to the programming model. 
To assist with our description, in Table~\ref{tab:modelsQuantum} we provide representative examples for each layer, which we elaborate  in subsequent subsections.

\begin{figure}
    \centering
    \includegraphics[width=0.7\columnwidth]{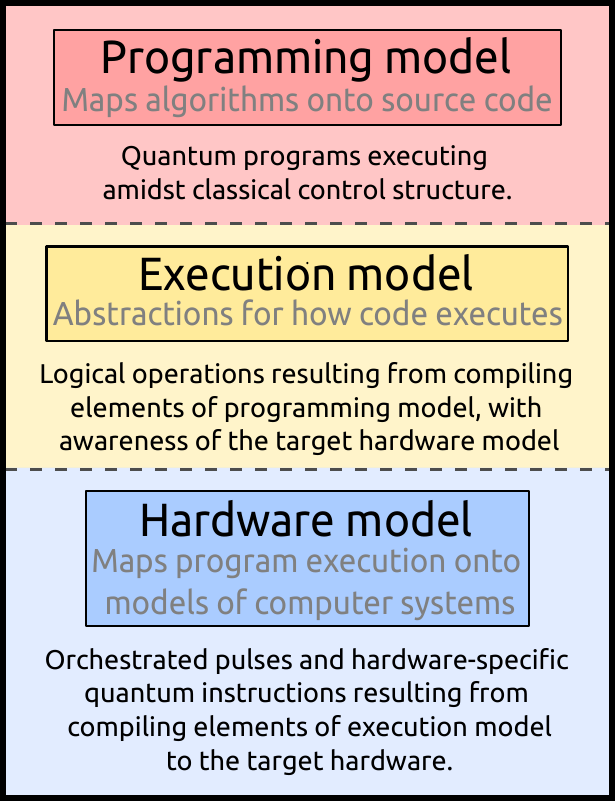}
    \caption{A proposed abstraction hierarchy for  quantum programming derived from the three-level model shown in Table \ref{tab:modelsExample}. While this idealized version depicts three separate layers, knowledge about adjacent layers is typically required in order to make good decisions. Furthermore, the state of today's quantum computing hardware and software can lead to overlapping, blurry boundaries, as will be seen in the examples in Section~\ref{sec:case-studies}. We depict this possibility by using permeable boundaries (dashed lines) between layers.}
    \label{fig:quantum-abstract-hierarchy}
\end{figure}

\begin{table}[htbp]
\caption{Examples of the three-model hierarchy applied to different quantum programming structures.
}
\begin{center}
\begin{tabular}{|l|l|}
\hline & Examples \\ \hline
\textbf{Programming model} & High-level quantum subroutines\\
 & Ansatz with hybrid optimization loop \\
  &  Unitary operations \\
  &  Repeat-until-success \\
\hline
\textbf{Execution Model} & Logical (non-hardware-native) circuits   \\
& Error mitigation \\
& Error correction \\
& QRASP \\
\hline
\textbf{Hardware model} & Hardware-native circuits  \\
& Pulses \\
\hline
\end{tabular}
\label{tab:modelsQuantum}
\end{center}
\end{table}%

While it preserves the simplicity and elegance of the abstraction hierarchy given in Table \ref{tab:modelsExample}, quantum computing today exposes a more complex picture derived from theoretical and engineering challenges. A significant amount of information flow between the models is required, and  there may even be \emph{leakage}, as will be exemplified in the case studies of Section~\ref{sec:case-studies}.
We chose to represent this with permeable boundaries (dashed lines) in Figure~\ref{fig:quantum-abstract-hierarchy}. For instance, the error-correcting code we choose will depend on the  hardware architecture. The way logical circuits are compiled and optimized also depends on the hardware model and available operations. The algorithmic primitives we use at the level of the programming model depend on our execution model (e.g., error correction vs. mitigation), and hardware model as well.

At present, two main hardware operation modalities dominate the quantum computing landscape: analog and digital. We prioritize here reasoning about abstraction hierarchies defined for digital, gate-based computing\footnote{In terms of expressiveness, the main challenge of analog quantum programming resides in the large number of details that permeate upwards from the hardware model directly into the execution and programming models. In consequence, analog quantum computing has a thin abstraction hierarchy and will not be the focus of our attention.} for a variety of reasons. First and foremost, the quantum circuit model allows re-use of concepts and methods from classical computing, thanks to evident similarities, as classical and quantum circuits map to well known logic forms that can be exploited effectively. It provides a more direct path to algorithmic reasoning, and already contains abstract machine models closer to those found in classical computing. Quantum algorithms written for digital quantum computers exhibit early signs of program composability. Finally, the path from pulses to native gates in hardware to quantum circuits with broader gate sets already provides some separation of concerns (albeit an imperfect one) which has triggered substantial innovation during the last decade.

\subsection{The  quantum hardware model}

The hardware model is an abstract definition of a particular quantum processor: the available operations (native gates, or pulses), connectivity, perhaps even properties of particular qubits.
In the hardware model, \emph{qubits} generalize classical bits to represent states which evolve as a consequence of quantum operations or by being measured. Physical quantum states are short lived, which forces the hardware model to be augmented with dynamical control considerations not found in classical hardware to try to extend quantum states long enough to be useful for computation. Properties of quantum states and their transition functions become resources about which one can reason approximately using linear logic, a specialized form of classical logic in which formulas involved in proofs can only be used once \cite{buss1998introduction}.

\emph{Approximately} is the key word in the discussion of abstraction hierarchies. Ideally, qubits would behave in accordance with a stable probability distribution determined by states that can be prepared without error, and whose measurement does not suffer from deviations from expected values due to extraneous interference. Because qubits do suffer from the latter, we are now forced to define \emph{quantum hardware} depending on the degree of approximation to an ideally behaving system. At the bare-metal hardware level -- furthest away from ideal behaviors -- sits the physical system that takes in modulated analog signals  (i.e., \emph{pulses}) and yields a quantum state in response. In a productive quantum programming hierarchy, information at this level should remain opaque to those interested in either the execution of programs or their development. Quantum control, while essential to quantum hardware platforms in practice, should produce guarantees about the state and reliability of the hardware that are indirectly harnessed to render native gates correctly. Weaving quantum control concerns into the execution or programming models reduces code portability and increases code complexity, both of which translate into slower innovation across the quantum stack.

Going up the abstraction ladder,  quantum hardware operations (i.e., gates, measurement) for a given native gate set are implemented by sequences of pulses. 
Tools such as BQSKit \cite{younis2021berkeley} and Superstaq \cite{campbell2023superstaq} can extract performance effectively at the pulse level for the hardware model, but only for small-scale systems.
Further, it is possible that particular gate sets in present quantum platforms are circumstantial to a given state of quantum engineering, and thus any directly realizable gates belong to the quantum hardware model.

\subsection{The quantum execution model}

The role of any execution model in a computing system is to subordinate the details involved in the hardware model, which are needed to produce computation, to expose only essential operations in a convenient form. That subordination is achieved through  an \emph{instruction}, an entity that serves both as an encoding of machine steps with a predictable outcome and a mnemonic interface for humans to program; an instruction's interface provides only the abstract transformations it produces and its hardware guarantees. In a classical microprocessor architecture, executing an instruction triggers a sequence of events across components in its internal organization; i.e., its division into modular components each with a single task.

In stark contrast to the quantum execution model, instructions in a classical computer do not make reference to individual gates. Bit-wise logic operations are available (e.g., \texttt{c = a \& b}), but programmers do not need to mentally joust with individual gate placement. We think  simultaneously about higher-order values (e.g., integers, floating-point) at our convenience. This results from operations in finite models of fixed-width arithmetic having the same algebraic structure as operations on the set of natural or real numbers. The number of bits is hence not fundamental, but circumstantial.

We argue that the core of the quantum execution model is logical circuits, wherein the instructions correspond to logical gates (i.e., non-hardware-native) and measurements. In an idealized case, a single logical circuit is executed once and the result returned. In more realistic cases, the execution model may incorporate more complex processes. Error mitigation is one example: a single instruction may involve an entire sequence in which multiple circuits are constructed, executed, and their results post-processed. Error correction, at a high level, is also part of an execution model: the instructions pertain to the specific code being used, e.g., lattice surgery operations in the surface code. However, their implementation details depend strongly on the underlying hardware model.

The resulting instruction set is then augmented with \emph{control flow} instructions. Control flow provides selective execution of instructions contingent on the result of comparisons between pieces of data in an algorithm. In classical computing, the execution model provides the essentials needed to capture procedures in terms of sequences of instructions, selection and iteration, and the model itself is given by a hardware-independent (but \emph{hardware-implementable}) abstraction known as random-access stored-program  (RASP) machines \cite{kaphengst1959abstrakte,elgot1964random}, capable of fully expressing facts about functions over recursively enumerable sets. The corresponding abstraction is the quantum random-access stored program  (QRASP) machine \cite{wang2023quantum}.

In a very concrete way, QRASPs are hardware-agnostic specifications for quantum circuit simulators, which happen to be implementable in actual quantum hardware platforms. By this token also, the primary abstraction provided by the execution model is that of unitary gates of arbitrary size measurementsand composition, a fundamental ingredient to achieve program modularity at the next level. However, contrary to the classical RASP machine, the existing QRASP model fails to provide productive insights into which high-level, quantum-native procedural structures may exist beyond classical ones; this is suspicious given the vastly larger resources in  quantum compared to classical computers.

\subsection{The quantum programming model}

To quote from above, the quantum programming model "maps the mental representation of a problem onto algorithms expressed in code." This could take many different forms, depending on the algorithm being applied. At a high level, we can see this as quantum programs \cite{ying2016foundations}, i.e., the composition of quantum subroutines, and their integration within classical control structure. However, it may still include individual gates, which we view as leakage into the execution model.

In our view, the responsibility of a programming model is to facilitate the translation of algorithms into executable code while preserving their intuitive meaning. We find here another instance of separation of concerns, eloquently captured by  Abelson: \textit{``programs must be written for people to read, and only incidentally for machines to execute''} \cite{abelson1996structure}. A programming model provides concepts and constructs to express a collection of problems conveniently. We call the resulting degree of convenience across a wide variety of tasks \emph{expressiveness}. Programming languages instantiate programming models, each of which attempts to attain a specific balance between  the programmer's required awareness of the underlying abstract machine that describes the execution model and the proximity to formally expressed facts and relations in a given problem specification.

Despite bearing a deeper abstraction hierarchy,  digital quantum programming is  somewhat similar  to its analog counterpart: the same (ideal) mathematical entities that describe hardware states are the entities we find in quantum programs. Current quantum-software abstractions try to alleviate the consequences of  preserving this conceptual complexity -- instead of simplifying it as we expect going  up the abstraction hierarchy -- by defining and using  high-level building blocks that combine quantum and classical objects, and drawing lines that mark conceptual and pragmatic boundaries between modules and kernels. Variational quantum algorithms  illustrate  the above. Yet, even with recent progress of quantum libraries, writing a quantum program remains closer to specifying hardware modules to be synthesized for a hypothetical quantum field-programmable gate array (FPGA) than to programming.

We argue, as a consequence of all the above, that quantum computing requires a distinct and better programming model above the circuit level.
Automatic oracle synthesis exemplifies this: it is convenient to synthesize functions as oracles, but it does not necessarily provide a new perspective for writing better quantum programs. Adding data types to quantum programs does not seem to suffice, since the process results in quantum composite types \cite{fu2020linear} instead of new atomic types with which one can reason easily about the behavior and correctness of a program without the need to run it. Quantum programming languages such as Quipper \cite{green2013quipper}, Silq \cite{bichsel2020silq}, Cavy \cite{mcnally2021practical}, Qrisp \cite{seidel2022qrisp} and Rhyme \cite{varga2024quantum} constitute some of the best ongoing explorations we know of that seek to understand how classical types (and analogous, native quantum types) might be interpreted in quantum programs and their structures. 

Based on the above, we also hypothesize that the limited number of fundamental quantum algorithms available so far may be explained by the lack of a higher-level programming model.  As a community, it is worth taking a direction similar to the one in classical computing which produced the famous equation \emph{algorithms + data structures = programs} \cite{wirth1976algorithms}.

\section{Case study in quantum programming: eigenvalue estimation}
\label{sec:case-studies}

In this section, we map a problem onto our abstraction hierarchy in two different ways. We choose eigenvalue estimation as our representative problem as it can be solved in multiple ways, most notably, (1) using a variational algorithm such as the VQE \cite{Peruzzo_2014}, and (2) using quantum phase estimation (QPE). While this initially may appear to be a comparison of NISQ vs. fault-tolerant approaches, we note it is  more nuanced, as one could, in principle, run a VQE on a fault-tolerant device with full error correction (or QPE on a NISQ processor), but most would agree this is not a judicious use of resources.

The result of our exercise is presented in Figure~\ref{fig:case-study-vqe} for VQE, and Figure~\ref{fig:case-study-qpe} for QPE. One notices immediately from these figures how partitioning of the workflow into models enables separation of concerns, but also how the barriers remain permeable. 
The models presented here are by no means exhaustive – these are just examples we think many readers will be familiar with, which also suggest that such a framework might be quite useful. There will inevitably be variations in implementation, and for this reason, we use pseudocode and try to keep the discussion high-level. In that spirit, we emphasize the purpose of these examples is to show how our model enables us to \emph{think and reason about the software stack}, not to provide hard-and-fast rules on what each layer contains.  

\subsection{VQE with error mitigation example}

In the VQE, the programming model consists primarily of:
\begin{itemize}
 \item designing an ansatz in terms of gates or higher-level subroutines,
 \item running a classical loop that optimizes its parameters. 
\end{itemize}

\begin{figure*}
    \centering
    \includegraphics[width=0.95\textwidth]{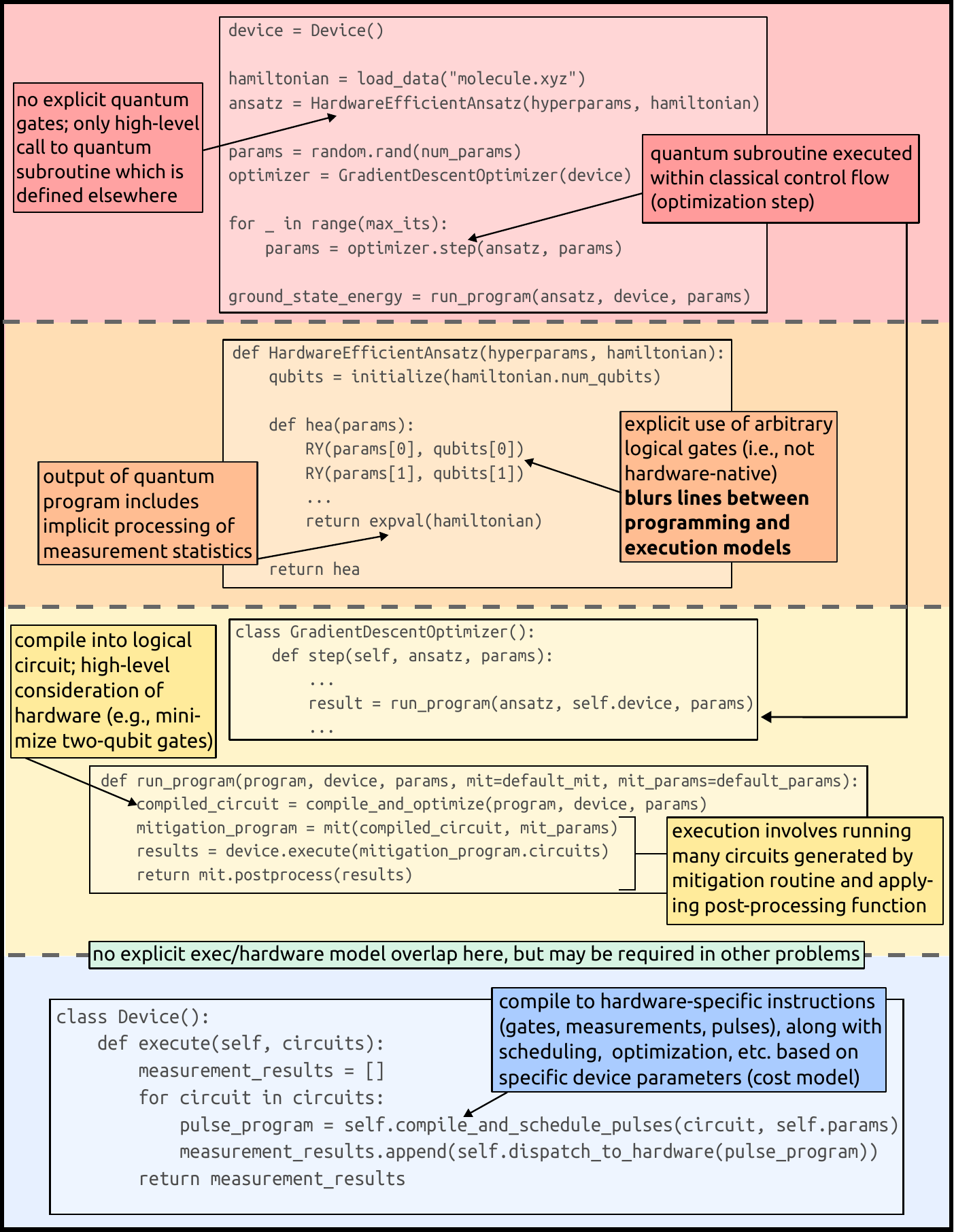}
    \caption{Mapping the eigenvalue estimation problem to our abstraction hierarchy using a variational algorithm like the variational eigensolver with
    error mitigation. The inset text describes information that flows between the layers. There is also leakage between the programming and execution models, shown in the overlapping orange box.
    }
    \label{fig:case-study-vqe}
\end{figure*}

\noindent These components together meet the definition of programming model from Sec \ref{ssec:parallel_computing}: "Maps the mental representation of a problem onto algorithms expressed in code". Ideally, at this level the user just needs to specify some high-level building blocks, and call a function that executes code on a quantum device under the hood. However, as we elaborate  below, there is both information flow and leakage into the execution model. 
 
The execution model comprises \emph{logical operations} that reflect how we envision the algorithm executing on hardware. This definition may initially seem at odds with the typical view of algorithms like VQE that run without error correction. Here, we take the view that "logical operation" refers to, e.g., gates applied in an ansatz without consideration for the particular hardware (e.g., Fermionic excitation gates). In our VQE the execution model contains two components that describe how the code executes:
\begin{itemize}
    \item an optimization step, where multiple circuits might be executed to compute expectation values or gradients,
    \item the entire algorithm running with error mitigation, which also modifies the "base circuits" coming from the ansatz.
\end{itemize}

Further to the second point, the user's mental picture in this model is that it is not just a single circuit that executes, but rather multiple circuits, whose results are post-processed (e.g., for the different noise scaling steps of a procedure like zero-noise extrapolation). This is hidden in the programming model, where a user only implements a parameter update step. 

There is nevertheless a certain amount of blurring. For instance, an ansatz may be described using a mix of individual logical gates and subroutines. Elements such as grouping strategy of Hamiltonian terms, or shot allocation strategy, are things a programmer must consider at the top level, but nevertheless pertain to the specifics of how code executes. For all these components, the routines used in the programming model must be synthesized into logical circuits, with awareness of any limitations in the hardware model (e.g., optimizing for 2-qubit gate count).

The responsibility of the hardware model is to translate a "logical circuit" into one that can execute on a particular QPU. This includes compilation to native gates and qubit topology, as well as scheduling and orchestrating gate execution. One could also consider compilation of logical operations straight into hardware-native pulses, or even designing an ansatz in terms of pulses \cite{liang2024napa}. However, we think it should be considered as a separate hardware model from the one presented in Figure~\ref{fig:case-study-vqe}\footnote{In particular, our reference to ``pulses'' as the most basic hardware operations is just  shorthand, as there exist hardware modalities which do not use pulses (e.g. photonics).}. 

While in Figure~\ref{fig:case-study-vqe} the execution and hardware models are distinct, there are techniques that blur the boundaries. For instance, error mitigation with dynamical decoupling is something we can picture as part of the execution model, however it is so intimately related to the hardware model that it is not immediately clear where it belongs.

To sum it up, the main abstractions at each level are:
\begin{itemize}
    \item Programming model: "VQA framework", i.e., ansatz, optimizer, etc.
    \item Execution model: "logical circuits", i.e., circuits which are not yet transpiled to hardware-native gate set; error mitigation method.
    \item Hardware model: "physical circuits", i.e., circuits within constraints of native gate set and connectivity.
\end{itemize}

\subsection{QPE with error correction}

\begin{figure*}
    \centering
    \includegraphics[width=0.95\textwidth]{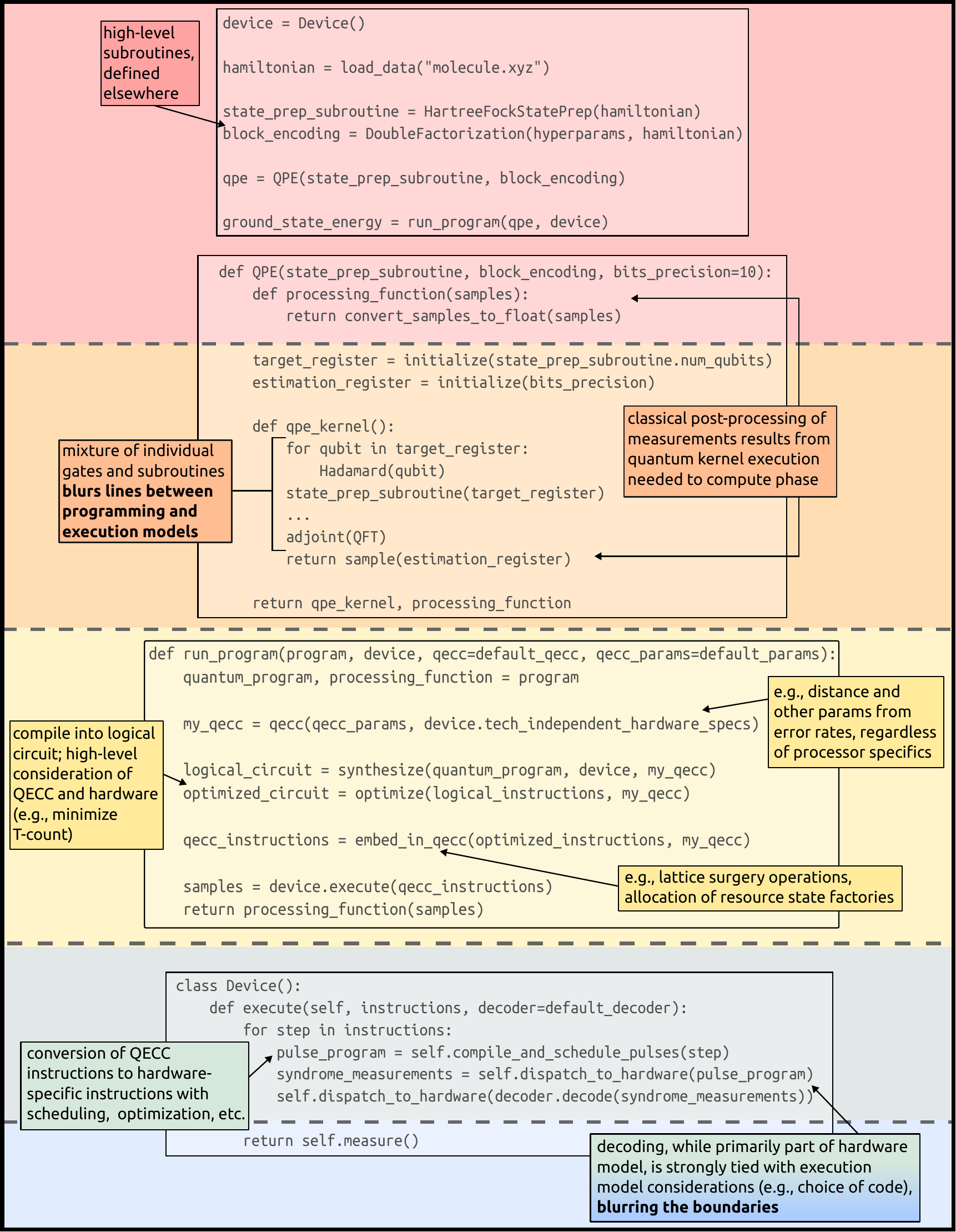}
    \caption{Mapping the eigenvalue estimation problem to our abstraction hierarchy using quantum phase estimation with full error correction. In addition to leakage between the programming and execution models, it becomes challenging to decouple the execution and hardware models when performing error-corrected quantum computing. While one can, for the most part, express everything with logical operations, this is complicated by hardware-adjacent tasks such as decoding.}
    \label{fig:case-study-qpe}
\end{figure*}

Quantum algorithms, like their classical counterparts, can be expressed as programs that represent a sequence of operations applied to a quantum system. One manifestation of this idea is  \emph{quantum While} programs \cite{ying2016foundations}. They present a grammar for quantum programs where possible operations include initializing qubits; applying unitary operations; applying a composition of programs; measuring; and applying programs based on post-selection of measurement outcomes. As a less formal definition, we refer to a more generic \emph{quantum subroutines model}, where quantum subroutines at varying levels of abstraction are composed to express an algorithm.

At a very low level of abstraction, a subroutine may correspond explicitly to a logical circuit. For instance, we can express the first step of QPE as a For loop that applies a Hadamard to each qubit (as portrayed in Figure~\ref{fig:case-study-qpe}). However, we can describe this more succinctly as a subroutine, \texttt{uniform\_superposition()}, without specifying how it is implemented. More complex algorithms will involve the nesting of many subroutines, among other operations, which leads to varying levels of blurriness between programming and execution models. 

In this case, the execution model starts at the level of logical circuits and gates, and contains most aspects of error correction. Optimization of this circuit in order to minimize certain resources (usually counts of qubits and T-gates) would also happen here. 
Despite the prevalence of circuits, one can imagine alternative models, for instance one using graph state formalism \cite{KrishnanVijayan_2024}, where it could be possible to compile higher-level subroutines directly into graph states without the need of using circuits as intermediate step.

Once this is done, logical operations get transformed into an intermediate representation that enables error correction (e.g., lattice surgery operations). 
Every logical operation will consist of (potentially thousands of) physical operations that constitute the native operations of the code, e.g., transversal gates, or full lattice surgery operations. 

Running in a full error-correcting code involves a significantly more complex interplay between the execution and hardware models; this is depicted by the two models overlapping in the green box in Figure~\ref{fig:case-study-qpe}. In particular, we must consider the interplay between quantum and classical devices required for decoding. For example, when using surface code, when a cycle of syndrome measurements is performed, the results are passed to a decoder, which decides which corrective operations should be applied. It is thus challenging to fully decouple the execution and hardware models. 

The hardware model includes:

\begin{itemize}
    \item set of elementary operations available for given hardware, 
    \item the architecture (e.g., baseline architecture or active-volume architecture as described in \cite{litinski2022active}),
    \item the decoding scheme.
\end{itemize}

The hardware model also includes translation of the error-correction operations to the hardware-native operations, which must then be scheduled and managed (and eventually lowered to pulses). This process will be extremely complex as it requires working with thousands or millions of physical qubits. A clear separation of concerns is essential here: someone working on logical operations in the execution model relies on them being correctly embedded into the code; but more importantly, those working on the error correction instructions must trust that all of this detailed scheduling happens effectively under the hood.

To summarize, the main abstractions for this example are:
\begin{itemize}
    \item Programming model: composition of quantum subroutines.
    \item Execution model: logical circuits; logical operations of the error correcting code.
    \item Hardware model: physical operations; architecture; decoding (coupled with execution model).
\end{itemize}

\section{Discussion}
\label{sec:discussion}

The state of software for quantum computing has emerged from the fast-paced race to morph experimental  devices into usable quantum computers. The speed at which quantum hardware and software platforms have evolved and matured into publicly usable devices is outstanding. This rate of change, however, is not sustainable as hardware becomes more complex en route to full fault-tolerant systems. Over time, the  number of qubits  will continue to grow, as will the number of details programmers need to grapple with. This complexity filters up from the hardware and impacts application software. Our task is to manage that complexity so hardware changes support the needs of software  without substantial disruptions to existing applications; i.e., letting software evolve without the need to rethink hardware.  To help us accomplish these goals, we need a layered hierarchy of abstractions corresponding to the levels of software development.

Sections~\ref{sec:quantum-abstraction-hierarchy} and~\ref{sec:case-studies} suggest that we have not yet arrived at a good abstraction hierarchy with clean interfaces between hardware, execution, and programming models. A better abstraction hierarchy would further facilitate separation of concerns between the layers. Figure \ref{fig:case-study-qpe} is telling of the prevailing mixture across levels. Hardware changes continue to impact quantum software, forcing programmers to be overly concerned with the details of the hardware as they develop code. We recognize that the concrete examples described in Section~\ref{sec:case-studies} do not resolve our many criticisms of the current state of affairs discussed in Section~\ref{sec:quantum-abstraction-hierarchy}. This means the abstraction hierarchy proposed in Section~\ref{sec:quantum-abstraction-hierarchy} is best seen as an intermediate step we can build on as we converge towards a consensus abstraction hierarchy.

A key indication of our work is the need for new quantum abstract machines that overcome the limitations of the QRASP model; we also note the severe limitations imposed by not having quantum randomly accessible memory (RAM) in our systems.  The circuit model on its own does not provide a clear path toward a solution. To reiterate our point above: an abstract machine that makes reference to gates and vector states provides an excellent basis to ground quantum hardware simulators, but not for a computing machine centered on algorithmic design and actual problem solving. Making headway into this facet of quantum computing will require a concerted (and well funded) effort of the quantum community supported by a systematic, empirical exploration of quantum programs \cite{nunez2023quapl} to detect the common patterns.

Achieving a good quantum abstraction hierarchy is a hard problem with a high payoff. This will require progress across multiple research vectors: 

\begin{itemize}
    \item \emph{Isolation of layers in the stack.} A good abstraction hierarchy supports a separation of concerns between layers. This would help the development communities at each level work concurrently, and reduce R\&D risks and costs. 
    \item \emph{Higher-level abstractions for improved productivity and exploration.} Abstractions for application developers should match cognitive representations of solutions, not details of the hardware. This would lead to more productive programming  and would let developers explore a wider range of algorithm ideas. 
    \item \emph{Formal software engineering practices for quantum computing.} A good abstraction hierarchy will help us develop testing and analysis tools to support best practices in software engineering \cite{zhao2020quantum,piattini2022quantum,murillo2024challenges}. This will reduce waste of limited quantum computer resources through broader correctness guarantees and runtime checks on hardware states. It will also support well-known principles in organization theory such as division of labor and rationalization of resources to quantitatively characterize and organize human processes.
\end{itemize}

All of this work would benefit from  knowledge built across several decades of
classical programming applied to quantum computing. The result would be an
increase in the pace of development of quantum computing and hopefully
avoidance of past mistakes.

\section{Conclusion} 

Quantum computing is relatively young.  Software engineering, however, is a mature discipline.  This paper is part of a long-term effort to bring established software engineering practices to quantum computing.  Drawing from experiences in parallel computing, which faced a similar problem 45 years ago, we assert that a key step is to define an abstraction hierarchy that supports a separation of concerns between the different facets of building software systems.   For example, people writing applications should not need to understand how compilers are constructed and compiler writers shouldn't need to be hardware engineers.

The parallel computing community over time settled on a three-layer abstraction hierarchy around which to organize their software development infrastructure.  We believe that the same hierarchy should be used for quantum computing:  a \textit{programming model} is used to map algorithms onto software, an \textit{execution model} to understand how software executes, and  a \textit{hardware model} to define an abstract machine that represents physical hardware.  This leads to the desired separation of concerns essential for any software engineering practice. Better quantum programming models have the advantage of being self-catalytic: good programs lead to better hardware, which result in even better programs.

To be clear, there is much work yet to be done.  What we accomplished in this paper is a proposed abstraction hierarchy validated by mapping known practices onto it.  This paper is a call to action for systematic work on abstraction hierarchies, not a presentation of fully fleshed-out and complete models.  Our hope is to drive a discussion in the quantum computing research community, reach consensus on the right abstractions, and then do the hard work of defining the right models for each layer in the hierarchy. This will create a conceptual foundation around which we will support highly productive software engineering for quantum computers. 

\section*{Acknowledgements} 

The authors thank Tom Lubinski and Yuval Sanders for useful discussions. ODM acknowledges funding from Canada's NSERC, the Canada Research Chairs program, and UBC. SNC acknowledges support and partial funding by the National Center for Supercomputing Applications and the Illinois Quantum Information Science and Technology Center, University of Illinois Urbana-Champaign.

\bibliographystyle{IEEEtran}  
\bibliography{references}

\end{document}